\documentclass[12pt]{article}
\input epsf

\newcommand{\news}{\setcounter{equation}{0}}
\newcommand{\be}{\begin{equation}}
\newcommand{\ee}{\end{equation}}
\newcommand{\bea}{\begin{eqnarray}}
\newcommand{\eea}{\end{eqnarray}}
\newcommand{\bean}{\begin{eqnarray*}}
\newcommand{\eean}{\end{eqnarray*}}
\font\upright=cmu10 scaled\magstep1
\font\sans=cmss12
\newcommand{\ssf}{\sans}
\newcommand{\stroke}{\vrule height8pt width0.4pt depth-0.1pt}
\newcommand{\Z}{\hbox{\upright\rlap{\ssf Z}\kern 2.7pt {\ssf Z}}}

\newcommand{\C}{{\rlap{\rlap{C}\kern 3.8pt\stroke}\phantom{C}}}
\newcommand{\R}{\hbox{\upright\rlap{I}\kern 1.7pt R}}
\newcommand{\CP}{\C{\upright\rlap{I}\kern 1.5pt P}}
\newcommand{\identity}{{\upright\rlap{1}\kern 2.0pt 1}}
\newcommand{\half}{\frac{1}{2}}
\newcommand{\pr}{\partial}

\begin{document}
\pagestyle{plain}

\title{\vskip -70pt
\begin{flushright}
{\normalsize DAMTP-2005-21} \\
\end{flushright}
\vskip 50pt
{\bf \Large \bf Superevolution}
 \vskip 30pt
}
\author{Nicholas S. Manton\thanks{email N.S.Manton@damtp.cam.ac.uk} \\[15pt]
{\normalsize
{\sl Department of Applied Mathematics and Theoretical Physics,}}\\
{\normalsize {\sl University of Cambridge,}}\\
{\normalsize {\sl Wilberforce Road, Cambridge CB3 0WA, England.}}\\
}
\vskip 20pt
\date{March 2005}
\maketitle
\vskip 20pt

\begin{abstract}
Usually, in supersymmetric theories, it is assumed that the
time-evolution of states is determined by the Hamiltonian, through
the Schr\"odinger equation. Here we explore the superevolution of
states in superspace, in which the supercharges are the principal 
operators. The superevolution equation is consistent with the Schr\"odinger
equation, but it avoids the usual degeneracy between bosonic and 
fermionic states. We discuss superevolution in supersymmetric
quantum mechanics and in a simple supersymmetric field theory.
\end{abstract}
\newpage
\section{Introduction}
\news
Supersymmetry is a beautiful symmetry, extending the Poincar\'e
symmetry of space-time. The supersymmetry generators are spinorial in
character, and they relate bosonic states with integer spin to fermionic
states of half-integer spin. Supersymmetry is present in a wide range
of theoretical settings -- quantum field theory, supergravity,
superstring theory. 
Yet there is no evidence so far of the physical realisation of 
supersymmetry. Even if supersymmetry were established through the discovery 
of heavier partner particles of the elementary particles we know, it 
would still not be exact supersymmetry but only an approximate, broken form.

A simpler version of supersymmetry occurs in supersymmetric quantum 
mechanics (see \cite{CGK} for a recent discussion of the various
formulations). Here, the states that are related by
supersymmetry may or may not differ in spin. Some rather special
systems with exact or approximate supersymmetry, in this more limited 
sense, are physically
realised. Certain nuclei have a sequence of excited states labelled by
increasing energy and angular momentum. The energy levels of one
nucleus with half-integer angular momenta have approximately the same 
spacing as the combined energy levels of
two close-by nuclei with integer angular momenta. These states can be
modelled using a supersymmetry algebra \cite{A}. Another example is an electron
in a magnetic field, restricted to a two-dimensional plane \cite{AC,CFKS}. The
magnetic field need not be uniform. Here, the states of the electron
with spin up are paired by a supersymmetry operator to states of the
same energy with spin down. Only the zero energy states are unpaired. 
(In this, and some other quantum mechanical examples, the states related by
supersymmetry are not really distinguished by the dichotomy
``bosonic/fermionic'', but it is convenient to use this terminology,
and we will do so in what follows.) In both
these examples, the states related by supersymmetry are physically
distinct, but have the same energy.

Yet there are many physical systems where there is a hint of
supersymmetry, without the accompanying energy degeneracy of states.
For example, part of a supersymmetric system can be something well-known, and
physically realised. A nice example has been explored by Kirchberg et
al. \cite{K}. This is the supersymmetric quantum mechanical Coulomb
system in ordinary three-dimensional space. The superpotential which
gives the Coulomb potential is quite simple, being $\half\lambda r$,
with $r$ the distance to the source and $\lambda$ a measure of its 
strength. In the formalism of Witten \cite{W}, applied to this example, 
the wavefunctions are differential
forms in $\R^3$, and the Hamiltonian commutes with the degree of the form. So
the Hilbert space splits up into the subspaces of 0-forms, 1-forms,
2-forms and 3-forms. Acting on 0-forms, the Hamiltonian is $-\nabla^2
+ \frac{\lambda}{r} + \frac{\lambda^2}{4}$, which for $\lambda < 0$ is
essentially the
Hamiltonian of the standard hydrogen atom, but with energy levels shifted
so that the ground state energy is zero. It is tempting to say that
this latter Hamiltonian is supersymmetric because it is part of a
larger, truly supersymmetric system. The full system has further
Hamiltonians, acting on 1-forms, 2-forms and 3-forms. These are
physically meaningful (for example, the Hamiltonian on 3-forms is the
repulsive Coulomb Hamiltonian), but they do not occur simultaneously
with the standard Coulomb Hamiltonian in the hydrogen atom.
Another example of this type is the Planck oscillator. This is the
standard quantum harmonic oscillator but with its ground state energy shifted
to zero. The quantized electromagnetic field can be regarded as an
infinite set of such oscillators, labelled by momentum and
polarisation.  A photon is the first excited state of one of these 
oscillators. The Planck oscillator also occurs naturally in a
supersymmetric context, but is then accompanied by a second, fermionic 
oscillator which has no physical role in the theory of electromagnetic 
radiation. 

Our goal, in this paper, is to find a convincing 
reinterpretation of supersymmetric systems, which avoids the
degeneracy between bosonic and fermionic states. Our hope is to obtain
a new understanding of physical systems previously regarded as, say, the
bosonic part of a 
supersymmetric system. We are particularly interested in quantum field
theory examples, where we want to retain the advantages of
supersymmetry, but avoid the mass degeneracy of bosonic
and fermionic particles. However, before studying quantum field
theory, we shall explore some quantum mechanical models.

Let us recall the relationship between factorizable Hamiltonians
\cite{S,IH,CKS} and the structure of supersymmetric quantum mechanics in one 
space dimension \cite{N,W1,SvH}. 
Suppose the Hamiltonian $H_0$ of a quantum particle can be factorized as
\be
H_0 = A^{\dag} A \,.
\ee
Then there is a related Hamiltonian, namely $H_1 = AA^{\dag}$. $H_0$
and $H_1$ are called partner Hamiltonians. The standard example is 
where $A = \frac{1}{\sqrt{2}}(\frac{d}{dx} + W(x))$, 
and $A^{\dag} = \frac{1}{\sqrt{2}}(-\frac{d}{dx} + W(x))$, so
\be
2H_0 = -\frac{d^2}{dx^2} + W(x)^2 - \frac{dW(x)}{dx} \,, \quad   
2H_1 = -\frac{d^2}{dx^2} + W(x)^2 + \frac{dW(x)}{dx} \,.
\ee
$H_0$ and $H_1$ both have non-negative spectrum, and any positive eigenvalue 
of $H_0$ is also a positive eigenvalue of $H_1$. This is because if
$A^{\dag} A \psi = E\psi$, then $AA^{\dag} A \psi = EA\psi$. 
So if $\psi$ is 
an eigenfunction of $H_0$ with eigenvalue $E$, then $A\psi$ is an 
eigenfunction of $H_1$ with eigenvalue $E$. This argument breaks down if 
$A\psi=0$, but in that case $E=0$.

$H_0$ could easily be a physical Hamiltonian for a particle in one
dimension, and we learn that there is a related physical Hamiltonian
$H_1$ with almost the same spectrum. However, the system does not
simultaneously have both Hamiltonians. The Planck
oscillator example is $W(x) = x$, where
\be
2H_0 = -\frac{d^2}{dx^2} + x^2 - 1 \,, \quad   
2H_1 = -\frac{d^2}{dx^2} + x^2 + 1 \,.
\ee
$H_0$ has spectrum $0,1,2,\dots$, and $H_1$ has spectrum $1,2,\dots$.

In the supersymmetric quantum mechanical analogue of partner Hamiltonians,
the Hilbert space is of the form ${\cal H} = {\cal H}_b \oplus 
{\cal H}_f$, and the wavefunction is a pair of ordinary functions 
$\pmatrix{\psi_0 \cr \psi_1}$ where $\psi_0$ is interpreted as the
bosonic part and $\psi_1$ as the fermionic part. The Hamiltonian has the
diagonal form
\be
H = \pmatrix{H_0&0\cr 0&H_1} = \pmatrix{A^{\dag} A&0\cr 0&AA^{\dag}} \,.
\ee
There is a hermitian supersymmetry operator
\be
Q = \pmatrix{0&A^{\dag}\cr A&0}
\ee
and the grading (Witten) operator
\be
K = \pmatrix{1&0\cr 0&-1} \,.
\ee
The complete supersymmetry algebra is
\be
K^2 = 1 \,, \quad \{K,Q\} = 0 \,, \quad Q^2 = H \,,
\label{SUSYalg}
\ee
from which it follows that $Q$ commutes with $H$.
This system has the two Hamiltonians $H_0$ and $H_1$ acting on the
different sectors, and it has degenerate fermionic and bosonic states
(with $E \ne 0$), connected by the action of $Q$. 

So far we have mentioned only Hamiltonians and their spectra, but
quantum mechanics is about the time evolution of states via the
Schr\"odinger equation, so let us look at this. In supersymmetric 
quantum mechanics, the usual Schr\"odinger equation is
\be
\pmatrix{i\frac{\pr \psi_0}{\pr t} \cr i\frac{\pr \psi_1}{\pr t}} = 
\pmatrix{A^{\dag} A&0\cr 0&AA^{\dag}} \pmatrix{\psi_0\cr \psi_1} \,,
\ee
so $\psi_0$ and $\psi_1$ evolve according to their respective
Hamiltonians. A key point is that $\psi_0$ and $\psi_1$ at some
initial time, say $t=0$, are independent. This leads to the
degeneracy between fermionic and bosonic states. $Q$ maps bosonic into
fermionic states, and vice versa, but plays no direct role in the 
time evolution.

One could impose a superselection rule forbidding linear
superpositions of fermionic and bosonic states. Even so, the initial
state could be either fermionic or bosonic, and there would still be two
independent states with the same energy.

An idea we have considered, but which is not our final proposal, is 
to restrict further, and impose the condition that
$\psi_1=0$ at the initial time. $\psi_1$ would then be zero for all time. 
In this way one would recover the Schr\"odinger equation for one
of the partner Hamiltonians, and there would be no degeneracy. The
ordinary quantum mechanics with Hamiltonian $H_0$ and wavefunction
$\psi_0$ becomes, in this way, part of a larger supersymmetric 
structure. One could simply pronounce that this restricted
Schr\"odinger equation is supersymmetric. This idea does not
properly exploit the supersymmetry, because there is no
explicit role for the supercharge $Q$. 

However, there is another route which captures the supersymmetric spirit of the
problem, leading to a very similar outcome. This involves an evolution 
equation in a superextension of physical time, and $\psi_1$ is 
non-vanishing. We describe this next. 

\section{Superevolution in Supersymmetric Quantum Mechanics}
\news

It has been known for a long time \cite{FW} that the Schr\"odinger equation of 
supersymmetric quantum mechanics has a ``square root'' in which the
evolution is determined by the supercharge $Q$. Let the operators be
as before. The superevolution equation is
\be
\pmatrix{i\frac{\pr \psi_0}{\pr t}\cr \psi_1} = 
\pmatrix{0&A^{\dag}\cr A&0} \pmatrix{\psi_0\cr \psi_1} \,,
\label{superevol}
\ee
which can be written more compactly as
\be
\pmatrix{i\frac{\pr}{\pr t}&-A^{\dag}\cr -A&1} 
\pmatrix{\psi_0\cr \psi_1}=0 \,.
\ee
The equations for the components are
\be
i\frac{\pr \psi_0}{\pr t} = A^{\dag}\psi_1 
\label{psi0eq}
\ee
\be
\psi_1 = A \psi_0 \,. 
\label{psi1eq}
\ee

Substituting (\ref{psi1eq}) into (\ref{psi0eq}) we see that $\psi_0$ obeys its 
Schr\"odinger equation, $i\frac{\pr \psi_0}{\pr t} =
A^{\dag}A\psi_0$. Taking the time derivative of (\ref{psi1eq}) and
substituting (\ref{psi0eq}) we see that $\psi_1$ obeys its Schr\"odinger
equation $i\frac{\pr \psi_1}{\pr t} =
AA^{\dag}\psi_1$. However, $\psi_0$ and $\psi_1$ are not independent. Indeed,
the general solution of the superevolution equation is obtained by
taking a solution of the Schr\"odinger equation for $\psi_0$ and then
setting $\psi_1 = A \psi_0$. We may regard $\psi_1$ as a shadow of the
physical state $\psi_0$. 

In some ways, postulating the superevolution
equation is hardly different from the earlier idea of just taking
$\psi_0$ evolving with its corresponding Hamiltonian. But it
has a more supersymmetric flavour, and still achieves the
desired result of avoiding the degeneracy of fermionic and bosonic
states. For each energy eigenstate of $H_0$ there is just one solution
of (\ref{superevol}), up to an overall normalisation constant.

Apart from the lack of independence of $\psi_0$ and $\psi_1$, there is
another significant difference between the superevolution
equation and the separate Schr\"odinger equations for $\psi_0$ and
$\psi_1$. This concerns the zero energy states. First, suppose that
$H_0$ has an eigenfunction $\phi$ with zero eigenvalue. Then $0 =
\langle \phi | H_0 | \phi \rangle = \langle \phi | A^{\dag} A | \phi \rangle = 
\langle A\phi | A \phi \rangle$, so $A \phi = 0$. Therefore, the corresponding
solution of (\ref{superevol}) is $\psi_0 = \phi$, $\psi_1 = 0$. 
Consistent with zero energy, there is no time dependence. Second, suppose that
$H_1$ has an eigenfunction $\widetilde\phi$ with zero eigenvalue. Then 
$A^{\dag} \widetilde\phi = 0$. Equation (\ref{psi0eq}) is solved by setting
$\psi_1 = \widetilde\phi$ and $\psi_0$ to be any time-independent
function, but for (\ref{psi1eq}) to be
satisfied one requires that $\widetilde\phi = A\psi_0$. This cannot be
solved, since it implies $\langle \widetilde\phi | \widetilde\phi \rangle
= \langle A\psi_0 | \widetilde\phi \rangle 
= \langle \psi_0 | A^{\dag} \widetilde\phi \rangle = 0$, and hence 
$\widetilde\phi = 0$, a contradiction.
We conclude that zero energy states of $H_0$ have corresponding
solutions of the superevolution equation, but zero energy states of
$H_1$ do not.

There is a supertime formulation of the superevolution equation. Let us extend
the time line to a supertime $\R^{1|1}$ with coordinates
$(t,\tau)$. $\tau$ commutes with $t$ but is odd, and $\tau^2=0$. The
supertime evolution operator $D$ is defined to be
\be
D = \frac{\pr}{\pr \tau} - \tau i \frac{\pr}{\pr t} \,.
\label{superD}
\ee
This acts on a wavefunction $\Psi$ which is a function of $t,\tau$ and
$x$. $\Psi$ has the expansion
\be
\Psi = \Psi_0 + \tau\Psi_1
\label{superPsi}
\ee
where $\Psi_0 \in {\cal H}_b$ and $\Psi_1 \in {\cal H}_f$ depend only 
on $t$ and $x$. Therefore
\be
D\Psi = \left(\frac{\pr}{\pr \tau} - \tau i \frac{\pr}{\pr t}\right)
(\Psi_0 + \tau\Psi_1) = \Psi_1 - \tau i \frac{\pr \Psi_0}{\pr t} \,.
\label{superDPsi}
\ee

Let us now extend the earlier definition of $Q$, by linearity, to
$\Psi$, with $Q$ acting on $\Psi_0$ and $\Psi_1$ as $Q$ previously
acted on $\psi_0$ and $\psi_1$. $Q$ and $\tau$ are taken
to anticommute. Therefore
\be
Q\Psi = Q\Psi_0 - \tau Q\Psi_1 = A\Psi_0 - \tau A^{\dag}\Psi_1 \,,
\label{superQPsi}
\ee
with $A\Psi_0 \in {\cal H}_f$ and $A^{\dag}\Psi_1 \in {\cal H}_b$. 
The superevolution equation is taken as
\be
D\Psi = Q\Psi \,.
\label{superevol1}
\ee
Combining (\ref{superDPsi}) and (\ref{superQPsi}), we see that in components
\bea
i\frac{\pr \Psi_0}{\pr t} &=& A^{\dag}\Psi_1 \\
\Psi_1 &=& A \Psi_0 \,.
\eea

It is easy to verify, abstractly or by acting on $\Psi$, that 
\be
D^2 = -i\frac{\pr}{\pr t} \,,
\ee
so $D$ is the square root of (minus) the time evolution operator that occurs
in the Schr\"odinger equation. One can check directly by acting on 
$\Psi$ that $DQ + QD = 0$. The superevolution equation is therefore 
a consistent square root of the Schr\"odinger equation, because it implies 
that
\be
-i\frac{\pr \Psi}{\pr t} = D^2\Psi = DQ\Psi = -QD\Psi = -Q^2\Psi = -H\Psi \,.
\ee

We have made a notational distinction between $\psi_0$, $\psi_1$,
which are ordinary functions of $x$ and $t$, and $\Psi_0$, $\Psi_1$,
for the following reason. It is best to regard $\Psi_0$ as even and
$\Psi_1$ as odd. This can be made explicit by extending space, with
coordinate $x$, to a superspace $\R^{1|1}$ with coordinates 
$(x,\theta)$, where
$\theta^2 = 0$ and $\theta$ and $\tau$ anticommute. Then let $\Psi_0 =
\psi_0$ and $\Psi_1 = \theta \psi_1$. The total wavefunction $\Psi$ 
becomes the even expression
\be
\Psi = \psi_0 + \tau\theta\psi_1 \,.
\ee
The operator $D$ is as before, but $Q$ becomes
\be
Q = \theta A + \frac{\pr}{\pr \theta} A^{\dag} \,.
\ee
Notice that both $D\Psi$ and $Q\Psi$ are odd. 
The superevolution equation (\ref{superevol1}) takes the form
\be
\left(\frac{\pr}{\pr \tau} - \tau i \frac{\pr}{\pr t}\right)
(\psi_0 + \tau\theta\psi_1)
= \left(\theta A + \frac{\pr}{\pr \theta} A^{\dag}\right)(\psi_0 +
\tau\theta\psi_1) \,.
\ee
This simplifies to
\be
\theta\psi_1 - \tau i \frac{\pr\psi_0}{\pr t}
= \theta A\psi_0 - \tau A^{\dag}\psi_1 \,.
\ee
Comparing coefficients of $\theta$ and $\tau$, we recover the
component equations 
\bea
i\frac{\pr \psi_0}{\pr t} &=& A^{\dag}\psi_1 \\
\psi_1 &=& A \psi_0 \,,
\eea
as before.

The superevolution equation (\ref{superevol1}) was
considered by Friedan and Windey \cite{FW} in the context of a
spinning supersymmetric particle, whose supersymmetry charge $Q$ is the
Dirac operator. This they exploited to prove the Atiyah-Singer index 
theorem. The initial state at $t=\tau=0$ was taken to be a spatial 
delta function, with vanishing dependence on odd spatial coordinates
like $\theta$. In contrast, we would allow the initial state
$\psi_0(t=0)$ to be an arbitrary, ordinary function of $x$. More importantly,
our proposal is for the physical realisation of supersymmetric quantum
mechanics, and not just a mathematical application. 

Superevolution equations have also been considered by Rogers \cite{R}, who
constructed a path integral representation for the finite 
(euclidean) supertime evolution operator $e^{-Ht-Q\tau}$. 
The initial state at $t=0$, $\tau = 0$ (in the context of
1-dimensional supersymmetric quantum mechanics) was taken to be a general 
function of $x$ and $\theta$, $\Psi = \phi_0(x) + \theta \phi_1(x)$; however 
we would impose the restriction that $\Psi$ is even. 
The wavefunction at $\tau = 0$
is then purely bosonic, and the fermionic part occurs multiplied by
$\tau$. This avoids the degeneracy between bosonic and
fermionic states. The physical wavefunction at a later time can be 
identified with $\Psi(t, \tau =0)$.

In quantum mechanics, it is not just the evolution of the wavefunction
that is important. One must also consider observables and their
expectation values. In superevolution, we are regarding $\psi_0$ as the
physical wavefunction, and $\psi_1 = A\psi_0$ as its shadow.
We propose that an observable should be a hermitian operator
acting on $\psi_0$. Let us define a normalised wavefunction to be one
satisfying $\langle\psi_0 | \psi_0 \rangle = 1$. For such a
wavefunction, the expectation value of an observable $O$ is
$\langle\psi_0 | O | \psi_0 \rangle$.

If $\psi_0$ is normalised, then the shadow wavefunction satisfies the
normalisation  $\langle\psi_1 | \psi_1 \rangle = \langle A \psi_0 | 
A \psi_0 \rangle = \langle\psi_0 | H_0 | \psi_0 \rangle = E$, where
$E$ is the energy expectation value. There are also shadow observables
$\widetilde O$ acting on $\psi_1$, but these can be related to
standard observables. The observable $O$ related to $\widetilde O$ is
given by
\be
\langle\psi_0 | O | \psi_0 \rangle = 
\langle\psi_1 | {\widetilde O} | \psi_1 \rangle
\ee
so $O = A^{\dag} {\widetilde O} A$. The expectation value of
$\widetilde O$ is defined as
\be
\frac{\langle\psi_1 | {\widetilde O} | \psi_1 \rangle}
{\langle\psi_1 | \psi_1 \rangle} 
= \frac{1}{E} \langle\psi_0 | O | \psi_0 \rangle \,. 
\ee
For example, if $\widetilde O = 1$ then $O = H_0$ and the expectation
value of $\widetilde O$ is $1$. If $\widetilde O = H_1 = AA^{\dag}$ 
then $O = (H_0)^2$
and the expectation value is $E$. The shadow observables do not make
sense in a state with $E=0$, because $\psi_1$ then vanishes.

\section{Supersymmetry and Differential Forms}
\news

Witten \cite{W} has formulated a large class of supersymmetric quantum
mechanical models in which the wavefunction is a differential form on
some finite-dimensional Riemannian manifold $M$. Many 
examples of supersymmetric quantum mechanics, including those
discussed in Sections 1 and 2, are special cases.

The basic model just involves the
geometry of $M$. Let $M$ have dimension $n$ and let $\Omega^{\rm ev}$ 
($\Omega^{\rm odd}$) denote the space of forms of even (odd)
degree. The complete Hilbert space is ${\cal H} = {\cal H}_b \oplus 
{\cal H}_f$ where ${\cal H}_b = \Omega^{\rm ev}$ and ${\cal H}_f = 
\Omega^{\rm odd}$. A wavefunction is therefore a pair $\Psi = 
\pmatrix{\omega^{\rm ev} \cr \omega^{\rm odd}}$ where $\omega^{\rm ev}
\in \Omega^{\rm ev}$, $\omega^{\rm odd} \in \Omega^{\rm odd}$.

The supersymmetry operator $Q$ is constructed from the de Rham
exterior derivative $d$ and its adjoint $\delta$. ($\delta = *d*$
acting on $\Omega^{\rm ev}$ when $n$ is odd, and $\delta = -*d*$ otherwise. 
$*$ is the Hodge duality operator, whose definition requires a 
Riemannian metric.) $d$ increases the degree of a form by 1, and $\delta$
decreases the degree by 1, so both operators map even forms to odd
forms, and vice versa. $d$ and $\delta$ have the properties $d^2 = 0$
and $\delta^2 = 0$, so $(d +\delta)^2 = d\delta + \delta d$, the
Laplace-Beltrami operator acting on forms on $M$.
The supersymmetry operators are  
\be
Q = \frac{1}{\sqrt{2}}\pmatrix{0&d +\delta \cr d +\delta&0}
\ee
\be
H = \half\pmatrix{d\delta + \delta d&0\cr 0&d\delta + \delta d} \,,
\ee
with $K$ as before. These satisfy the algebra (\ref{SUSYalg}).
Notice that, formally, $Q$ and $H$ act in the same way on 
$\Omega^{\rm ev}$ and $\Omega^{\rm odd}$, but this is rather an
illusion, since the detailed formulae depend on the degree.

A key observation of Witten is that the operators above
can be modified to include a real-valued, superpotential function $h$ defined 
on $M$. One simply replaces $d$ by $d_h = e^{-h} d \, e^h$ and $\delta$
by $\delta_h = e^h \delta \, e^{-h}$ in $Q$ and $H$. (Note that this
is not a trivial conjugation, as it would be if $\delta_h$ were $e^{-h} \delta
\, e^h$.) The algebra (\ref{SUSYalg}) is still satisfied.

The component equations for superevolution in Witten's model are
\be
i \frac{\pr}{\pr t} \omega^{\rm ev} = \frac{1}{\sqrt{2}} 
(d_h + \delta_h) \omega^{\rm odd} 
\label{omegaeveq}
\ee
\be
\omega^{\rm odd} = \frac{1}{\sqrt{2}} 
(d_h + \delta_h) \omega^{\rm ev} \,.
\label{omegaoddeq}
\ee
This again avoids the degeneracy between bosonic and fermionic
states that occurs with the Schr\"odinger evolution, governed by the
Hamiltonian $H$. A special case reproduces the 1-dimensional 
supersymmetric quantum mechanical model of Section 2. Choose $M = \R$ 
and set $W(x) = \frac{dh(x)}{dx}$. The wavefunction is the pair
\be
\Psi = \pmatrix{\psi_0 \cr \psi_1 \, dx}
\ee
where $\psi_0$ and $\psi_1$ are ordinary functions of $x$ and $t$. 
The abstract $\theta$ of Section 2 is here replaced by $dx$. 
Then (\ref{omegaeveq}) and (\ref{omegaoddeq}) reduce to
\bea
i \frac{\pr}{\pr t} \psi_0 &=& 
\frac{1}{\sqrt{2}}(d_h + \delta_h) \psi_1 \, dx \nonumber \\
&=& -\frac{1}{\sqrt{2}} (e^h *d* (e^{-h} \, \psi_1 \, dx)) \nonumber \\
&=& \frac{1}{\sqrt{2}}\left(-\frac{\pr \psi_1}{\pr x} +
\frac{dh}{dx}\psi_1 \right) \nonumber \\
&=& A^{\dag}\psi_1
\eea
and
\bea
\psi_1 \, dx &=& \frac{1}{\sqrt{2}}(d_h + \delta_h) \psi_0 \nonumber \\
&=& \frac{1}{\sqrt{2}} \, e^{-h} d(e^h \psi_0) \nonumber \\
&=& \frac{1}{\sqrt{2}}\left(\frac{\pr \psi_0}{\pr x} +
\frac{dh}{dx}\psi_0 \right) \, dx \nonumber \\
&=& (A\psi_0) \, dx \,.
\eea
which reproduce equations (\ref{psi0eq}) and (\ref{psi1eq}).

In Witten's model, the superspace version of the superevolution
equation can be expressed entirely in terms of differential forms, and
standard operations on forms. Recall that the fermionic wavefunction 
$\omega^{\rm odd}$ is odd as it stands. We identify
$\tau$ with the 1-form $dt$, and then extend Witten's formalism by 
making the wavefunction into a differential form on ${\widetilde M} = M \times 
[-\infty , \infty]$, where the second factor is the time axis. The 
wavefunction is now taken to be
\be
{\widetilde \Psi} = \omega^{\rm ev} + dt \wedge \omega^{\rm odd} \,,
\ee
an even form on ${\widetilde M}$. The operator 
$D = \frac{\pr}{\pr \tau} - \tau i \frac{\pr}{\pr t}$ becomes
\be
{\widetilde D} = \iota_{\frac{\pr}{\pr t}} - dt \, i \frac{\pr}{\pr t} \,,
\ee
where $\iota_{\frac{\pr}{\pr t}}$ is the inner product operator that
cancels a $dt$ factor immediately to the right (and gives zero acting
on a form with no $dt$ in it), and $dt$ acts by left exterior multiplication. 
$Q$ is replaced by the essentially identical ${\widetilde Q} = 
\frac{1}{\sqrt{2}}(d_h + \delta_h)$, which is defined to act just 
on the $M$ variables, and which anticommutes with $dt$.
In this extended formalism, the superevolution equation for Witten's
model becomes
\be
{\widetilde D}{\widetilde \Psi} = {\widetilde Q}{\widetilde \Psi} \,.
\ee
This reduces to the component equations (\ref{omegaeveq}) and 
(\ref{omegaoddeq}).
 
Notice that the Witten model is not relativistic. Even
if $M = \R^n$ and there is no superpotential, the superevolution 
equation is not Lorentz invariant in $\R^{n+1}$.

\section{Superevolution in Field Theory}
\news

In this section we consider the simplest supersymmetric quantum field 
theory in 1+1 dimensions, the theory of one real scalar field and one 
Majorana fermion field \cite{F}, and present its superevolution equations. But 
first, we present the conventional interpretation of the
field theory, and its Schr\"odinger equations.

It is standard to write down the Lagrangian first, and then
canonically quantize. However, the Majorana condition implies that the 
Majorana field is conjugate to itself, and this leads to some
ambiguities in factors of 2. This difficulty can be resolved using a
Dirac constraint formalism but we won't go through this. Instead, 
we shall simply
state the canonical commutation and anticommutation relations for the
field operators, and give the algebra of supersymmetry operators.

Let $x$ (or $y$) denote the spatial coordinate. The
scalar field $\phi(x)$ and its conjugate momentum $\pi(x)$ are
independent hermitian operators at each point. In the Schr\"odinger 
representation they have no time dependence. 

The Dirac matrices obey $(\gamma^0)^2
= 1$, $(\gamma^1)^2 = -1$ and $\gamma^0\gamma^1 + \gamma^1\gamma^0 =
0$. We shall use the Majorana representation for these:
\be
\gamma^0 = \pmatrix{0&i\cr -i&0} \,, \quad \gamma^1 =  
\pmatrix{0&i\cr i&0}  \,. 
\ee
The Majorana spinor field $\psi(x) = \pmatrix{\psi_1(x)\cr \psi_2(x)}$
has two components, both of which are hermitian operators.

The non-vanishing canonical commutation and anticommutation relations
are
\bea
[\phi(x), \pi(y)] &=& i\delta(x-y) \\
\{\psi_\alpha(x), \psi_\beta(y) \} &=& \delta_{\alpha\beta}\delta(x-y) \,,
\eea
with all commutators $[\phi, \phi]$, $[\pi, \pi]$ and 
$[\phi, \psi_\alpha]$, $[\pi, \psi_\alpha]$ vanishing.

The Hamiltonian is
\be
H = \half\int\left( \pi^2 + (\pr_x \phi)^2 + W(\phi)^2 + i\psi_1
\pr_x\psi_1 - i\psi_2\pr_x\psi_2 
+ 2i\frac{dW(\phi)}{d\phi}\psi_1\psi_2 \right)\, dx \,,
\ee
where $W(\phi)$ is an arbitrary function of $\phi$, usually assumed to
be a polynomial. The total momentum operator is 
\footnote{Our signs are such that $H$ and $P$ are the time and space
components of the covariant 2-vector $P_{\mu}$.}
\be
P = \half\int\left( 2\pi \pr_x \phi + i\psi_1
\pr_x\psi_1 + i\psi_2\pr_x\psi_2 \right)\, dx \,.
\ee
It is best to combine these into the combinations
\be
H+P = \half\int\left( (\pi + \pr_x \phi)^2 + W(\phi)^2 
+ 2i\psi_1\pr_x\psi_1 + 2i\frac{dW(\phi)}{d\phi}\psi_1\psi_2 \right)\,
dx 
\label{HplusP}
\ee
\be
H-P =  \half\int\left( (\pi - \pr_x \phi)^2 + W(\phi)^2 
- 2i\psi_2\pr_x\psi_2 + 2i\frac{dW(\phi)}{d\phi}\psi_1\psi_2 \right)\,
dx \,.
\label{HminusP}
\ee
The two supercharges are
\be
Q_1 = \int\left( (\pi + \pr_x \phi)\psi_1 - W(\phi)\psi_2 \right)\, dx 
\label{SUcharge1}
\ee
\be
Q_2 = \int\left( (\pi - \pr_x \phi)\psi_2 
+ W(\phi)\psi_1 \right)\, dx \,.
\label{SUcharge2}
\ee
The theory simplifies to a free theory if $W(\phi) = m\phi$. The
particles associated with the quantized scalar and Majorana field then 
both have mass $m$. 

After a somewhat long calculation, using the canonical (anti)commutation 
relations, one can verify that the above operators obey the
supersymmetry algebra
\be
Q_1^2 = H+P 
\label{SUSY1}
\ee
\be
Q_2^2 = H-P
\label{SUSY2}
\ee
\be 
Q_1 Q_2 + Q_2 Q_1 = 0 \,.
\label{QAnticomm}
\ee
These imply that $Q_1$ and $Q_2$ commute with both $H$ and $P$. 
Formally, a boundary contribution appears as a central charge on the
right hand side of (\ref{QAnticomm}) but it vanishes if we suppose 
that $\phi$ has equal vacuum expectation values as $x \to \pm\infty$.

As an example of part of the calculation of $Q_1^2$, consider the square
of the term involving $\pi\psi_1$. Symmetrizing in the spatial
variables of integration $x$ and $y$ this becomes
\be
\half\int\int\left(\pi(x)\psi_1(x)\pi(y)\psi_1(y) 
+ \pi(y)\psi_1(y)\pi(x)\psi_1(x) \right) \, dx \, dy \,,
\ee
and since $\pi$ commutes with itself and with $\psi_1$ this simplifies
to
\bea
&&  \half\int\int \pi(x)\pi(y)(\psi_1(x)\psi_1(y) 
+ \psi_1(y)\psi_1(x)) \, dx \, dy \nonumber \\
&\quad&\quad\quad\quad = \half\int \pi(x)\pi(y) \delta(x-y) \, dx \, dy
\nonumber \\
&\quad&\quad\quad\quad = \half\int (\pi(x))^2 \, dx \,.
\eea

In the Schr\"odinger picture, states evolve in time according to the
Hamiltonian. Let $\Psi$ denote the complete quantum state, and $T$ the
time. $\Psi$ obeys the Schr\"odinger equation
\be
i\frac{\pr \Psi}{\pr T} = H \Psi \,.
\ee
If $\Psi$ is an eigenstate of $H$ with total energy $E$, then
\be
\Psi(T) = \Psi(0) e^{-iET} \,.
\ee
One normally regards $\Psi(0)$ as a superposition of states 
completely specified by the
numbers and momenta of the various particles in the theory. For
example, in the free theory, for one scalar particle of mass $m$ and 
momentum $p$, the energy is
$E_p = \sqrt{p^2 + m^2}$, and one would write $\Psi(0) =
|p\rangle$ and $\Psi(T) = |p\rangle e^{-iE_p T}$.

However, for our purposes this is inadequate, because it does not give
a satisfactory representation of the total momentum and spatial displacement
operators. We need to regard states as functions of time, $T$, and of
the spatial centre of mass coordinate $X$. A general state is written
as $\Psi(T,X)$. The state $\Psi(0,0)$ is a linear
combination of the multiparticle states $|p_1,p_2,\dots,p_n\rangle$,
where for each momentum $p_i$ we need to specify the particle type too. 

We can now impose more symmetrical time and space evolution equations
on the state $\Psi$, namely
\be
i\frac{\pr \Psi}{\pr T} = H \Psi 
\label{Schrospace}
\ee
\be
i\frac{\pr \Psi}{\pr X} = P \Psi \,.
\label{Schrotime}
\ee
If, as usual, $\Psi$ is an eigenstate of both $H$ and $P$, with
the eigenvalues $E$ and $P'$ being the total energy and momentum, then
\be
\Psi(T,X) = \Psi(0,0) e^{-i(ET + P'X)} \,.
\ee

This phase dependence on the location of the centre of mass is standard
in the quantum mechanics of one or more particles \cite{D}, but in 
field theory it is generally neglected. It is implicit in field theory, as one
always regards $e^{-iPa}$ as the operator that spatially displaces a state by
$a$. With our notation we have, explicitly,
$e^{-iPa} \Psi(T,X) = \Psi(T,X+a)$.
\footnote{Another way in which the dependence is implicit in field theory is
that the relationship between the particle creation operator
$a_p^{\dag}$ and the field operators $\phi(x)$ and $\pi(x)$ involves
$e^{ipx}$, and this changes phase if one displaces the spatial
origin. So a one-particle state of momentum $p$ changes under a
displacement.}

We can regard the pair of equations (\ref{Schrospace}) and 
(\ref{Schrotime}) as the Schr\"odinger equations of quantum field 
theory in 1+1 dimensions. In our supersymmetric field theory,
the Hilbert space of states decomposes as ${\cal H} = {\cal H}_b \oplus 
{\cal H}_f$, where states in ${\cal H}_b$ have even fermion number,
and states in ${\cal H}_f$ have odd fermion number.  
$\Psi$ can be a general element of ${\cal H}$, but normally one
imposes the superselection rule that $\Psi$ is either in ${\cal H}_b$
or in ${\cal H}_f$. The action of $Q_1$ and
$Q_2$ maps states in ${\cal H}_b$ to physically distinct states 
in ${\cal H}_f$, and vice versa, degenerate in both energy and momentum.

This completes our summary of the standard interpretation of the field theory.

Now we show that, because of the supersymmetry, 
the Schr\"odinger equations of the field theory can be replaced by
superevolution equations. To do this we extend (1+1)-dimensional
spacetime to a superspace $\R^{2|2}$ with coordinates 
$T, X, \theta_1, \theta_2$,
where $\theta_1$ and $\theta_2$ are odd. We introduce a state
in superspace $\Psi(T,X,\theta_1,\theta_2)$ and consider its
expansion in $\theta_1$ and $\theta_2$,
\be
\Psi(T,X,\theta_1,\theta_2) = \Psi_0(T,X)
+ \theta_1\Psi_1(T,X) + \theta_2\Psi_2(T,X)
+ \theta_1\theta_2\Psi_{12}(T,X) \,.
\ee
By analogy with what we did in supersymmetric quantum mechanics, we
require that $\Psi_0$ and $\Psi_{12}$ lie in ${\cal H}_b$, and 
$\Psi_1$ and $\Psi_2$ lie in ${\cal H}_f$. We also treat $\Psi_1$ and
$\Psi_2$ as odd, anticommuting with $\theta_1$ and $\theta_2$.

Our assumptions mean that the expansion of $\Psi$ is analogous to that of
a superfield in $\R^{2|2}$ whose
components are classical bosonic and fermionic fields; however here
the components of $\Psi$ are multi-particle quantum states.

The superspace evolution operators are
\be
D_1 = \frac{\pr}{\pr \theta_1} - \theta_1 i \pr_+
\ee
and
\be
D_2 = \frac{\pr}{\pr \theta_2} - \theta_2 i \pr_- \,,
\ee
where $\pr_{\pm} = \pr_T \pm \pr_X$. They obey the relations
\bea
D_1^2 &=& -i\pr_+ \\
D_2^2 &=& -i\pr_- \\
D_1D_2 + D_2D_1 &=& 0 \,.
\eea
Because of the close formal similarity of this algebra with the
supersymmetry algebra (\ref{SUSY1}) -- (\ref{QAnticomm}), we can impose 
the consistent pair of superevolution equations
\be
D_1 \Psi = Q_1 \Psi 
\label{superfieldevol1}
\ee
\be
D_2 \Psi = Q_2 \Psi \,.
\label{superfieldevol2}
\ee
By acting with $D_1$ and $D_2$ on each of these equations, and noting
that $D_1$ and $D_2$ anticommute with $Q_1$ and $Q_2$, we
verify that these superevolution equations imply the Schr\"odinger
equations (equivalent to (\ref{Schrospace}) and (\ref{Schrotime}))
\be
i\pr_+ \Psi = (H + P)\Psi 
\label{Schro+}
\ee
\be
i\pr_- \Psi = (H - P)\Psi \,.
\label{Schro-}
\ee

It is worthwhile to expand both eqs.(\ref{superfieldevol1}) and 
(\ref{superfieldevol2}) in their components, to check their
consistency. The first equation gives
\bea
\Psi_1 &=& Q_1 \Psi_0 \\
i\pr_+ \Psi_0 &=& Q_1 \Psi_1 \\
\Psi_{12} &=& -Q_1 \Psi_2 \\
i\pr_+ \Psi_2 &=& -Q_1 \Psi_{12} \,.
\eea
Since $Q_1^2 = H + P$, we can verify that each component state
$\Psi_0$, $\Psi_1$, $\Psi_2$ and $\Psi_{12}$ obeys
(\ref{Schro+}). Similarly the second superevolution equation gives
\bea
\Psi_2 &=& Q_2 \Psi_0 \\
i\pr_- \Psi_0 &=& Q_2 \Psi_2 \\
\Psi_{12} &=& Q_2 \Psi_1 \\
i\pr_- \Psi_1 &=& Q_2 \Psi_{12} \,,
\eea
which implies that each component obeys (\ref{Schro-}). From both sets of
equations together, we see that the components are related
algebraically to $\Psi_0$ by
\be
(\Psi_0, \, \Psi_1, \, \Psi_2, \, \Psi_{12}) = (\Psi_0, \, Q_1\Psi_0, 
\, Q_2\Psi_0, \, -Q_1 Q_2\Psi_0) \,.
\ee
Provided $\Psi_1$, $\Psi_2$ and $\Psi_{12}$ are related to $\Psi_0$ in
this way, and provided $\Psi_0$ obeys the pair of Schr\"odinger
equations (\ref{Schro+}) and (\ref{Schro-}), it follows that all the component 
equations above are satisfied.

So, as in supersymmetric quantum mechanics, the only independent state
is $\Psi_0$, which is in the bosonic part of the Hilbert space, 
${\cal H}_b$. $\Psi_1$, $\Psi_2$ and $\Psi_{12}$ are
shadow states that accompany $\Psi_0$ in the superevolution, but they
carry no independent physical information. There is no physically 
independent fermionic state obtained by the action of $Q_1$ or
$Q_2$ on a bosonic state.

We shall explore below, in a little more detail, the superevolution of
particle states in free field theory. 

\section{Free Field Theory}
\news

The free theory of one scalar and one Majorana field, both of mass
$m$, is diagonalised
by passing to momentum space. One can directly see the particle
content, and can clarify the physics of the superevolution equations.

The scalar field operators $\pi(x)$ and $\phi(x)$ have the coupled
momentum space expansions
\bea
\pi(x) &=& \int \frac{dp}{2\pi} (-i) \sqrt{{\frac{E_p}{2}}}
(c_p e^{-ipx} - c_p^{\dag} e^{ipx}) \\ 
\phi(x) &=& \int \frac{dp}{2\pi}  \frac{1}{\sqrt{2E_p}}
(c_p e^{-ipx} + c_p^{\dag} e^{ipx}) 
\eea
where $E_p = \sqrt{p^2 + m^2}$.\footnote{$(E_p, p)$ are the components of a
covariant 2-vector $p_{\mu}$.} The canonical commutation relations require
\bea
&\quad& [c_p, c_{p'}^{\dag}] = 2\pi\delta(p-p') \\
&& [c_p, c_{p'}] = [c_p^{\dag}, c_{p'}^{\dag}] = 0 \,.
\eea

For the Majorana field $\psi_\alpha(x)$ we first need to present the
solutions of the classical Dirac equation
\be
(i \gamma^\mu \pr_\mu - m)\psi = 0 \,.
\ee
In the Majorana representation, this becomes
\bea
\pr_+ \psi_2 &=& -m\psi_1 \\
\pr_- \psi_1 &=& m\psi_2 \,.
\eea
Plane wave solutions of positive frequency (energy) are of the form
\be
\psi(t,x) = u(p) e^{-i(E_p t + px)}
\ee
where 
\be
u(p) = \pmatrix{\sqrt{E_p + p}\cr -i\sqrt{E_p - p}} \,.
\ee
Similarly, there are negative frequency plane wave solutions
\be
\psi(t,x) = v(p) e^{i(E_p t + px)}
\ee
with 
\be
v(p) = \pmatrix{\sqrt{E_p + p}\cr i\sqrt{E_p - p}} \,.
\ee
The momentum space expansion of the Majorana field 
$\psi = \pmatrix{\psi_1 \cr \psi_2}$ is
\be
\psi(x) = \int\frac{dp}{2\pi}  \frac{1}{\sqrt{2E_p}} (a_p u(p)e^{-ipx}
+ a_p^{\dag} v(p)e^{ipx}) \,.
\ee
The expressions for $u(p)$ and $v(p)$ imply that both components
of $\psi$ are hermitian. To satisfy the canonical anticommutation
relations one requires
\bea
&\quad& \{ a_p, a_{p'}^{\dag}\} = 2\pi\delta(p-p') \\
&& \{ a_p, a_{p'}\} = \{ a_p^{\dag}, a_{p'}^{\dag}\} = 0 \,.
\eea
One also requires that the operators $c$, $c^{\dag}$ commute with
$a$, $a^{\dag}$.

Starting from the formulae (\ref{HplusP}) -- (\ref{SUcharge2}), 
with $W(\phi) = m\phi$, and performing a number of integrations, one 
can obtain the momentum space expressions for $H$, $P$,
$Q_1$ and $Q_2$. These can be normal ordered without
discarding infinite constants, because of the supersymmetry, and one
finds
\be
H \pm P = \int\frac{dp}{2\pi}(E_p \pm p)(c_p^{\dag}c_p +
a_p^{\dag}a_p) 
\label{HPmom}
\ee
\be
Q_1 = i\int\frac{dp}{2\pi}\sqrt{E_p + p} \, (-a_p^{\dag}c_p + c_p^{\dag}a_p)
\label{Q1mom}
\ee
\be
Q_2 = \int\frac{dp}{2\pi}\sqrt{E_p - p} \, (a_p^{\dag}c_p + c_p^{\dag}a_p) \,.
\label{Q2mom}
\ee

The (Fock) vacuum $|0\rangle$ is annihilated by $c_p$ and $a_p$ for
all $p$. It is therefore annihilated by all the operators 
$Q_1$, $Q_2$, $H$ and $P$, so it is supersymmetric and has zero energy
and momentum.
 
Let us define bosonic and fermionic 1-particle states by
\be
|p_b\rangle =  c_p^{\dag} \, |0\rangle \,, \quad 
|p_f\rangle =  a_p^{\dag} \, |0\rangle \,.
\ee
(We ignore the additional factors of $\sqrt{2E_p}$ required for a
relativistic normalisation.)
There is just one solution of the superevolution equations that can be
constructed from these. $\Psi_0$ should be bosonic, so, at
$(T,X)=(0,0)$,  we set it equal
to $|p_b\rangle$. Then, by acting with $Q_1$ and $Q_2$, as given by
(\ref{Q1mom}) and (\ref{Q2mom}), we find that
\be
\Psi_1 = -i\sqrt{E_p + p} \, |p_f\rangle \,, \quad 
\Psi_2 = \sqrt{E_p - p} \, |p_f\rangle \,, \quad
\Psi_{12} = -im \, |p_b\rangle \,.
\ee
We must multiply all these component states by $e^{-(E_p T + pX)}$ to 
obtain their values at $(T,X)$. Superevolution implies that the only physical
particle is a scalar boson, although its shadow states $\Psi_1$ and
$\Psi_2$ are fermionic.

2-particle states can be constructed similarly. The obvious 2-boson
state is $\Psi_0 = c_{p}^{\dag}c_{p'}^{\dag} \, |0\rangle$, with energy
$E = E_p + E_{p'}$ and momentum $P' = p + p'$. This generates a
solution of the superevolution equations with the shadow states
\bea
\Psi_1 &=& -i\sqrt{E_p + p} \, a_{p}^{\dag}c_{p'}^{\dag} \, |0\rangle
- i\sqrt{E_{p'} + p'} \, a_{p'}^{\dag}c_{p}^{\dag} \, |0\rangle \\
\Psi_2 &=& \sqrt{E_p - p} \, a_{p}^{\dag}c_{p'}^{\dag} \, |0\rangle
+ \sqrt{E_{p'} - p'} \, a_{p'}^{\dag}c_{p}^{\dag} \, |0\rangle \\
\Psi_{12} &=& -i\left(\sqrt{(E_{p'} + p')(E_p - p)} 
- \sqrt{(E_p + p)(E_{p'} - p')}\right) a_{p}^{\dag}a_{p'}^{\dag} \, |0\rangle
\nonumber \\
&& \quad\quad\quad\quad -2im c_{p}^{\dag}c_{p'}^{\dag}|0\rangle \,.
\eea
All these must be multiplied by $e^{-(E T + P' X)}$. 
The coefficient of the first term in $\Psi_{12}$ simplifies to
$i\sqrt{2(E_pE_{p'} - pp' - m^2)}$, where the positive square root is
taken if $p > p'$, and the negative root if $p < p'$. A further
simplification is possible by introducing a rapidity variable
$\lambda$, such that $E_p = m\cosh\lambda$ and $p = m\sinh\lambda$, and
similarly $\lambda'$. Then
this coefficient becomes $2im\sinh{\half(\lambda - \lambda')}$.
Notice that not only $\Psi_0$, but also all the shadow states are
symmetric under the interchange of $p$ and $p'$.

There is a further candidate state for $\Psi_0$, namely $\Psi_0 
= a_{p}^{\dag}a_{p'}^{\dag} \, |0\rangle$, which is also in ${\cal H}_b$, 
since it is a 2-fermion state. The shadow states associated with this
are rather similar to those given above. This state, and similar
multi-particle states with an even number of fermions, are the most
problematic for our superevolution proposal. We were hoping for an
interpretation of supersymmetric field theory with only one type of
particle. We have managed to exclude the 1-fermion state, but appear
to need to allow 2-fermion states. 

We have the following thoughts about this problem. First, note that
the state $a_{p}^{\dag}a_{p'}^{\dag} \, |0\rangle$ is not directly related
to $c_{p}^{\dag}c_{p'}^{\dag} \, |0\rangle$ by supersymmetry (although it
occurs in combination with $c_{p}^{\dag}c_{p'}^{\dag} \, |0\rangle$ in
$\Psi_{12}$ above), and it is
probably an accident of the free field theory that these two states
are degenerate in energy and momentum. In the interacting theory, the
2-boson sector and the 2-fermion sector may be physically quite
different, having different 2-particle to 2-particle scattering
amplitudes, and different bound states (if any). This would
follow from the different permutational symmetry. Both in the free and 
interacting theory, the 2-boson states are symmetric under particle 
interchange, and the 2-fermion states are antisymmetric. As a shortcut 
to ensure that there is only one type of physical particle, we could perhaps
require that all multi-particle states are totally symmetric. This
proposal is consistent with the superevolution equations, because the
action of $Q_1$ and $Q_2$ preserves the symmetry of states, but
whether it is consistent in the interacting theory requires further
investigation.

\section{Conclusion}
\news 

We have revived the idea that the fundamental evolution equation in 
supersymmetric quantum mechanics should be a ``square root'' of the
Schr\"odinger equation. This means treating the supersymmetry charge
as an evolution operator in a superspace, and we call the resulting
equations the superevolution equations. The supersymmetry algebra
implies that if the superevolution equations are satisfied then so is
the Schr\"odinger equation. Usually, in supersymmetric quantum
mechanics, there are degenerate bosonic and fermionic states which are
physically distinct, and linearly independent, but the superevolution equations
relate them, so the degeneracy disappears. One version of the 
superevolution takes place in a rather abstract superspace, but in
Witten's model of supersymmetric quantum mechanics, the superevolution
equations can be presented using standard techniques from the theory
of differential forms.

We have extended the notion of superevolution to a simple
supersymmetric field theory in 1+1 dimension. To make this work we
needed to clarify the idea that the Schr\"odinger equation in quantum field
theory determines the evolution of states in both time and space (via
the Hamiltonian and total momentum operators). The superevolution
equations use the supercharges to define a consistent 
superspace evolution. Again, the superevolution equations imply that
the Schr\"odinger equation is satisfied, but the space of solutions is
smaller, because the superevolution relates states that are usually
treated as physically independent. As a result there is a suppression
of the degeneracy between bosonic and fermionic 1-particle states. A
natural choice leads to a unique supersymmetric vacuum of zero energy
and momentum, and the only physical 1-particle state being bosonic. 2-particle 
bosonic states also occur, as desired, but it could be problematic to remove 
the 2-fermion states. We suggested a way to deal with these too, 
leading to a theory which retains its supersymmetric character, but
which has only bosonic physical particles.

One might ask, in this case, what the fermions are doing. They would
contribute internal lines to Feynman diagrams (the vertices are
determined by the interaction term
$2i\frac{dW(\phi)}{d\phi}\psi_1\psi_2$ of the Hamiltonian). The best
interpretation might be that the supersymmetric theory defines a
special way of quantizing the purely bosonic field theory, leading to
all the usual advantages of supersymmetry (finiteness, zero vacuum
energy), but without physical fermions. The fermions are then rather like
the ghosts that occur in gauge theories (but we prefer to call them shadows).

It is of course important to explore extensions of the
ideas here to higher dimensions, and to investigate whether it is possible
to have a supersymmetric interpretation of a theory with
just fermions, or of a theory like QED, which has spin 1 
photons and spin $\half$ (non-Majorana) electrons.

\section*{Acknowledgements}

I would like to thank Adam Ritz for a discussion, and Anne Schunck for
helpful comments.

\end{document}